\documentclass[prd,amsmath,amssymb,showpacs,superscriptaddress,nofootinbib,twocolumn]{revtex4-1}
\usepackage{multirow}
\usepackage{float}
\usepackage{graphicx}
\usepackage{dcolumn}
\usepackage{bm}
\usepackage{rotating}
\usepackage{color}
\usepackage{subfigure}
\usepackage{pstricks}
\usepackage{soul}
\usepackage{overpic}
\usepackage[english]{babel}
\usepackage[colorlinks,linkcolor=blue,anchorcolor=blue,citecolor=blue]{hyperref}
\usepackage{tabularx}
\usepackage{comment}
\usepackage{lineno}
\usepackage{threeparttable}

\newcommand{\psip}{\psi(3686)}
\newcommand{\psipp}{\psi(3770)}
\newcommand{\jpsi}{J/\psi}
\newcommand{\chicz}{\chi_{c0}}
\newcommand{\chico}{\chi_{c1}}
\newcommand{\chict}{\chi_{c2}}
\newcommand{\chicJ}{\chi_{cJ}}
\newcommand{\chicj}{\chi_{c0,2}}

\newcommand{\piz}{\pi^0}

\newcommand{\kpkm}{K^{+}K^{-}}

\newcommand{\GG}{\gamma\gamma}
\newcommand{\GGG}{\gamma\gamma(\gamma)}
\newcommand{\EE}{e^+e^-}

\newcommand{\mulC}{\multicolumn}

\addto{\captionsenglish}{

 }

\begin{document}

\title{\boldmath Improved measurements of two-photon widths of the $\chicJ$ states and helicity analysis for $\chict\to\GG$}

\author{
\begin{center}
\begin{small}
M.~Ablikim$^{1}$, M.~N.~Achasov$^{9,e}$, S. ~Ahmed$^{14}$, M.~Albrecht$^{4}$, A.~Amoroso$^{53A,53C}$, F.~F.~An$^{1}$, Q.~An$^{50,a}$, J.~Z.~Bai$^{1}$, Y.~Bai$^{39}$, O.~Bakina$^{24}$
% ...
\\
\vspace{0.2cm}
(BESIII Collaboration)\\
\vspace{0.2cm} {\it
$^{1}$ Institute of High Energy Physics, Beijing 100049, People's Republic of China\\
% ...
$^{4}$ Bochum Ruhr-University, D-44780 Bochum, Germany\\
% ...
$^{9}$ G.I. Budker Institute of Nuclear Physics SB RAS (BINP), Novosibirsk 630090, Russia\\
% ...
$^{14}$ Helmholtz Institute Mainz, Johann-Joachim-Becher-Weg 45, D-55099 Mainz, Germany\\
% ...
$^{24}$ Joint Institute for Nuclear Research, 141980 Dubna, Moscow region, Russia\\
% ...
$^{39}$ Southeast University, Nanjing 211100, People's Republic of China\\
% ...
$^{50}$ University of Science and Technology of China, Hefei 230026, People's Republic of China\\
% ...
$^{53}$ (A)University of Turin, I-10125, Turin, Italy; (B)University of Eastern Piedmont, I-15121, Alessandria, Italy; (C)INFN, I-10125, Turin, Italy\\
% ...
$^{a}$ Also at State Key Laboratory of Particle Detection and Electronics, Beijing 100049, Hefei 230026, People's Republic of China\\
% ...
$^{e}$ Also at the Novosibirsk State University, Novosibirsk, 630090, Russia\\
% ...
}
\vspace{0.4cm}
\end{small}
\end{center}
}

\date{\today}

\begin{abstract}

Based on $448.1 \times 10^6$ $\psip$ events collected with the BESIII detector, the decays $\psip\to\gamma\chicJ, \chicJ \to \gamma\gamma~(J=0, 1, 2)$ are studied. The decay branching fractions of $\chicj \to \gamma\gamma$ are measured to be $\mathcal{B}(\chicz\to\GG) = (1.93 \pm 0.08 \pm 0.05 \pm 0.05)\times 10^{-4}$ and $\mathcal{B}(\chict\to\GG) = (3.10 \pm 0.09 \pm 0.07 \pm 0.11)\times 10^{-4} $, which correspond to two-photon decay widths of $\Gamma_{\GG}(\chicz) = 2.03 \pm 0.08 \pm 0.06 \pm 0.13 ~\rm{keV}$ and $\Gamma_{\GG}(\chict) = 0.60 \pm 0.02 \pm 0.01 \pm 0.04 ~\rm{keV}$ with a ratio of $\mathcal{R}=\Gamma_{\GG}(\chict)/\Gamma_{\GG}(\chicz)= 0.295 \pm 0.014 \pm 0.007  \pm 0.027$, where the uncertainties are statistical, systematic and associated with the uncertainties of $\mathcal{B}(\psip\to\gamma\chicj)$ and the total widths $\Gamma(\chicj)$, respectively. For the forbidden decay of $\chico\to\GG$, no signal is observed, and an upper limit on the two-photon width is obtained to be $\Gamma_{\GG}(\chico)<5.3 ~\rm{eV}$ at the 90\% confidence level. The ratio of the two-photon widths between helicity-zero and helicity-two components in the decay $\chict\to\GG$ is also measured to be $f_{0/2} = \Gamma^{\lambda=0}_{\GG}(\chict)/\Gamma^{\lambda=2}_{\GG}(\chict) = (0.0 \pm 0.6 \pm 1.2)\times 10^{-2}$, where the uncertainties are statistical and systematic, respectively.
\end{abstract}

\maketitle

 \section{Introduction}

Charmonium physics is at the boundary between perturbative and non-perturbative quantum chromodynamics~(QCD). Notably, the two-photon decays of \textit{P}-wave charmonia are helpful for better understanding the nature of inter-quark forces and decay mechanisms~\cite{Quark01,Quark02}. In particular, the decays of $\chicj \to \GG$ offer the closest parallel between quantum electrodynamics~(QED) and QCD, being analogous to the decays of the corresponding triplet states of positronium. To the lowest order, for charmonium the ratio of the two-photon decay widths is predicted to be~\cite{TwoPhoton}
\begin{equation}
\mathcal{R}^{(0)}_\text{th} \equiv \frac{\Gamma(^{3}P_{2}\to\GG)}{\Gamma(^{3}P_{0}\to\GG)} = 4/15 \approx 0.27.
\end{equation}
Any discrepancy from this lowest order prediction can arise due to QCD radiative corrections or relativistic corrections. The measurement of $\mathcal{R}$ provides useful information on these effects. Theoretical predictions on the decay rates are obtained using a non-relativistic approximation~\cite{Pre01,Pre02}, potential model~\cite{Pre03}, relativistic quark model~\cite{Pre04,Pre05}, non-relativistic QCD factorization framework~\cite{Pre06,Pre11}, effective Lagrangian~\cite{Pre07}, as well as lattice calculations~\cite{Pre08}. The predictions for the ratio $\mathcal{R}\equiv\Gamma_{\GG}(\chict)/\Gamma_{\GG}(\chicz)$ cover a wide range of values between 0.09 and 0.36. The decay $\chico \to \GG$ is forbidden by the Landau-Yang theorem~\cite{Yang}. Precise measurements of these quantities will guide the development of theory.

The two-photon decay widths of $\chicj$ have been measured by many experiments~\cite{PDG}. Using the decay of $\psip\to\gamma\chicj, \chicj\to\GG$, both CLEO-c and BESIII experiments reported results of the two-photon decay widths $\Gamma_{\GG}(\chicj)$~\cite{CLEOc,LUOT}.  BESIII has now collected the largest $\psip$ data sample in $\EE$ collisions, which provides a good opportunity to update and improve these measurements.

Additionally, in the decay $\chict\to\GG$, there are two independent helicity amplitudes, $i.e.$, the helicity-two amplitude ~($\lambda =2$) and the helicity-zero amplitude~($\lambda=0$), where $\lambda$ is the difference between the helicity values of the two photons. The corresponding ratio between the two-photon partial widths of the two helicity components, $f_{0/2} = \Gamma^{\lambda=0}_{\GG}(\chict)/\Gamma^{\lambda=2}_{\GG}(\chict)$, is predicted to be less than 0.5$\%$~\cite{Pre02}, while the previous experimental results from BESIII~\cite{LUOT} is $f_{0/2} = (0 \pm 2 \pm 2)\times 10^{-2}$. A more precise measurement of this ratio can be used to test the QCD prediction.

In this paper, we perform an analysis of $\psip\to\gamma\chicJ, \chicJ \to \GG$ ~(throughout the text, $\chicJ$ presents $\chi_{c0,1,2}$ unless otherwise noted).  The decay branching fractions are measured and the corresponding two-photon decay width $\Gamma_{\GG}(\chi_{cJ})$ are extracted. We also determine the ratio of two-photon decay width~($\mathcal{R}$) between the $\chict$ and $\chicz$ as well as of the two helicity components in the $\chict\to\GG$, $f_{0/2}$.

\section{The BESIII Experiment and Data Set}

This analysis is based on a sample of $448.1 \times 10^6$ $\psip$ events~\cite{NpsipAll} collected with the BESIII detector~\cite{BES} operating at the BEPCII collider~\cite{BEPC}. In addition, the off-resonance data sample taken at $\sqrt{s}=3.65$~GeV, corresponding to an integrated luminosity of 48 $\rm{pb^{-1}}$ ~\cite{Offpsip}, and the $\psipp$ data sample taken at $\sqrt{s} = 3.773$~GeV, corresponding to an integrated luminosity of 2.93 $\rm{fb^{-1}}$~\cite{Psipp}, are used to study the continuum background.

The BESIII detector features a nearly cylindrically symmetry and covers 93\% of the solid angle around the $\EE$ interaction point~(IP). The components of the apparatus, ordered by distance from the IP, are a 43-layer small-cell main drift chamber~(MDC), a time-of-flight~(TOF) system based on plastic scintillators with two layers in the barrel region and one layer in the end-cap region, a 6240-cell CsI(Tl) crystal electromagnetic calorimeter~(EMC), a superconducting solenoid magnet providing a 1.0~T magnetic field aligned with the beam axis, and resistive-plate muon-counter layers interleaved with steel. The momentum resolution for charged tracks in the MDC is 0.5\% for a transverse momentum of 1~GeV/$c$. The energy resolution in the EMC is 2.5\% in the barrel region and 5.0\% in the end-cap region for 1~GeV photons. Particle identification~(PID) for charged tracks combines measurements of the energy loss, $dE/dx$, in the MDC and flight time in the TOF and calculates probabilities $\text{prob}(h)$ ($h=p, \pi, K$) for each hadron ($h$) hypothesis. More details about the BESIII detector are provided elsewhere~\cite{BES}.

The optimization of event selection criteria and the estimation of the physical backgrounds are performed using Monte Carlo~(MC) simulated samples. The GEANT4-based~\cite{GEANT4} simulation software BOOST~\cite{BOOST} includes the geometric and material description of the BESIII detectors, the detector response and digitization models, as well as the tracking of the detector running conditions and performance. The production of the $\psip$ resonance is simulated by the MC event generator KKMC~\cite{KKMC}, while its decays are generated by EVTGEN~\cite{EvtGen} for known decay modes with branching ratios being set to the world average values in Particle Data Group~(PDG)~\cite{PDG}, and by LUNDCHARM~\cite{LundCharm} for the remaining unknown decays. For the simulation of the continuum process, $\EE\to\GGG$, the Babayaga~\cite{Babayaga} QED event generator is used.

\section{Data Analysis}

\begin{figure}[htbp]
   \centering
  \begin{overpic}[width=7.5cm,height=6.0cm,angle=0]{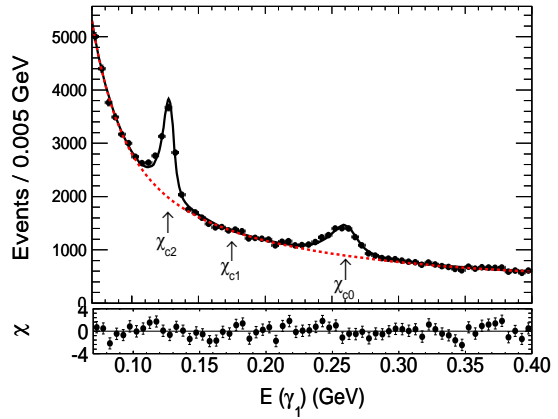}
   \end{overpic}
  \vskip -0.8cm
  \hskip 0.5cm
   \caption{(color online). Upper plot: The fitted $E(\gamma_1)$ spectrum for the $\psip$ data sample. The dots with error bar indicate data, the~(black) solid line is the best fit result, and the~(red) dashed line shows the background. The expected positions of the  $\chicz$, $\chico$, $\chict$ are indicated by arrows. Lower plot: The number of the standard deviations~($\chi$) of the data points for the best fit result. }  \label{FIT}
  \end{figure}

The event selection for the final states follows the same procedure as described in Ref.~\cite{LUOT}. It requires no charged tracks and three photon candidates, each with $E(\gamma)>70$ MeV and $|\cos \theta|<0.75$, where $E(\gamma)$ is the energy of the photon candidate, $\theta$ is the angle of the photon with respect to the positron beam direction. This requirement is used to suppress continuum background, $\EE\to\GGG$, where the two energetic photons have high probability of distributing in the forward and backward regions. The average interaction point of each run is assumed as the origin for the selected candidates. A four-constraint~(4C) kinematic fit is performed by constraining the total four momentum to that of the initial $\EE$ system, and events with $\chi^2_{4C} \le 80$ are retained. The energy spectrum of the radiative photon, $E(\gamma_1)$, which has the smallest energy among the three photon candidates, is shown in Fig.~\ref{FIT}, where structures associated with the $\chicz$ and $\chict$  are clearly observed over substantial backgrounds.

To determine the signal efficiencies, three signal MC samples, each with 1.2 million events, are generated by setting the mass and width of $\chicJ$ to the PDG values. For the radiative transition $\psip \to \gamma \chi_{c0,1}$, the angular distributions of the cascade \textit{E}1 transitions~\cite{E1} follow the formulae in Refs.~\cite{E11,E12}, and a uniform angular distribution is used to generate the process $\chi_{c0,1} \to \GG$. The full angular distribution used for $\psip \to \gamma\chict, \chict \to \GG$ is discussed in association with Eq.~(\ref{Chic2}) in Sec.~\ref{Hilicity}. The signal MC is generated with $\chict\to\GG$ in a pure helicity-two process, because the helicity-zero component is negligible relative to the helicity-two component as verified in Sec.~\ref{Hilicity}. For the \textit{E}1 transitions, the phase space is expected to have an energy dependence of $E_{\gamma}^3$, where $E_{\gamma}$ is the energy of the radiative photon in the center-of-mass system of the parent particle~\cite{Energy}.

The energy resolutions of the radiative photon are  $\sigma(E(\gamma_1)) = 5.91 \pm 0.05$~MeV for $\chicz$ and $\sigma(E(\gamma_1)) = 3.43 \pm 0.01 $~MeV for $\chict$, determined by the MC simulation. The efficiencies for the $\chicz$ and $\chict$ are $\epsilon(\chicz) = (40.88 \pm 0.04)\%$ and $\epsilon(\chict) = (39.85 \pm 0.04)\%$, respectively.

The dominant non-peaking background is from the continuum process $\EE\to\GGG$. MC simulations show that the backgrounds from $\psip$ radiative decays into $\eta$, $\eta^\prime$, and $3\gamma$ are non-peaking, spread over the full range of $E(\gamma_1)$, and the overall magnitude is less than 0.2\%. Therefore, these backgrounds do not significantly change the shape of the dominant continuum background and are neglected. In addition, we investigate possible sources of peaking backgrounds by using the inclusive $\psip$ MC sample. It is found that the process $\chicj\to\piz\piz(\eta\eta)$ with $\piz(\eta)\to\GG$ may produce a peak around the signal region, where two of the photons are not detected or are outside of the fiducial volume of the detector. We generate 100M events of each channel to determine the efficiencies of the peaking backgrounds; the expected numbers of peaking background are calculated by incorporating the decay branching fraction from Ref.~\cite{PDG} and are summarized in Table~\ref{ExMC01}.

\begin{table}[htpb]
\caption{Expected number of peaking background events in the $\chicj$ signal regions from MC simulation. The uncertainties are associated with the uncertainty of decay branching fractions in Ref~\cite{PDG}. }\label{ExMC01}
\begin {tabular}{lcr@{.}l@{$\pm$}r@{.}clcr@{.}l@{$\pm$}r@{.}cl}
\hline\hline
Decay Modes & \mulC{6}{c}{$n_{\chicz}$}& \mulC{6}{c}{$n_{\chict}$} \\
\hline
$\psip \to \gamma\chicj,\chicj\to\piz\piz$  & & 115& 8 & 10 & 2 & & & 27 & 0 &  2 & 5 &\\
$\psip \to \gamma\chicj,\chicj\to\eta\eta$ & & 5    & 3 &  0  & 5 & & &  1  & 0 &  0 & 1 & \\
\hline
Sum                                                          & & 121& 1 &10  & 2 & & & 28 & 0 & 2 & 5 &\\
\hline\hline
\end{tabular}
\end{table}

\section{Measurement of Branching Fractions and Two-photon Widths}

\begin{figure}[htbp]
   \centering
    \begin{overpic}[width=7.5cm,height=6.5cm,angle=0]{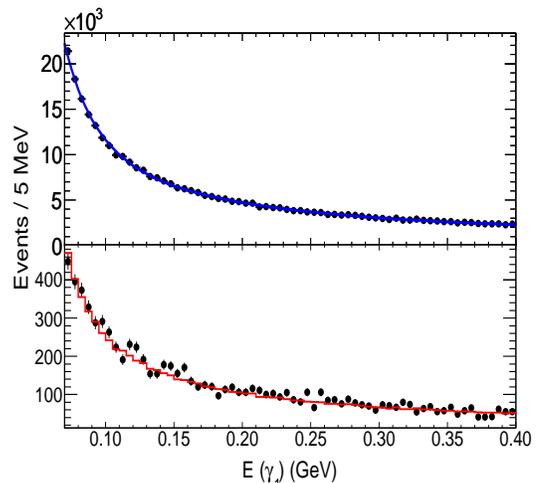}
   \end{overpic}
  \vskip -0.8cm
  \hskip 0.5cm
   \caption{(color online). Background $E(\gamma_1)$ spectrum. Upper plot: The best fit result~(blue solid line) to $\psipp$ data~(dots with error bar) using Eq.~(\ref{BKG}). Lower plot: The comparison of $E(\gamma_1)$ spectrum between off-$\psip$ data~(dots with error bar) and $\psipp$ data~(red histogram). }\label{background}
  \end{figure}

\begin{figure}[htbp]
   \centering
  \begin{overpic}[width=7.5cm,height=5.0cm,angle=0]{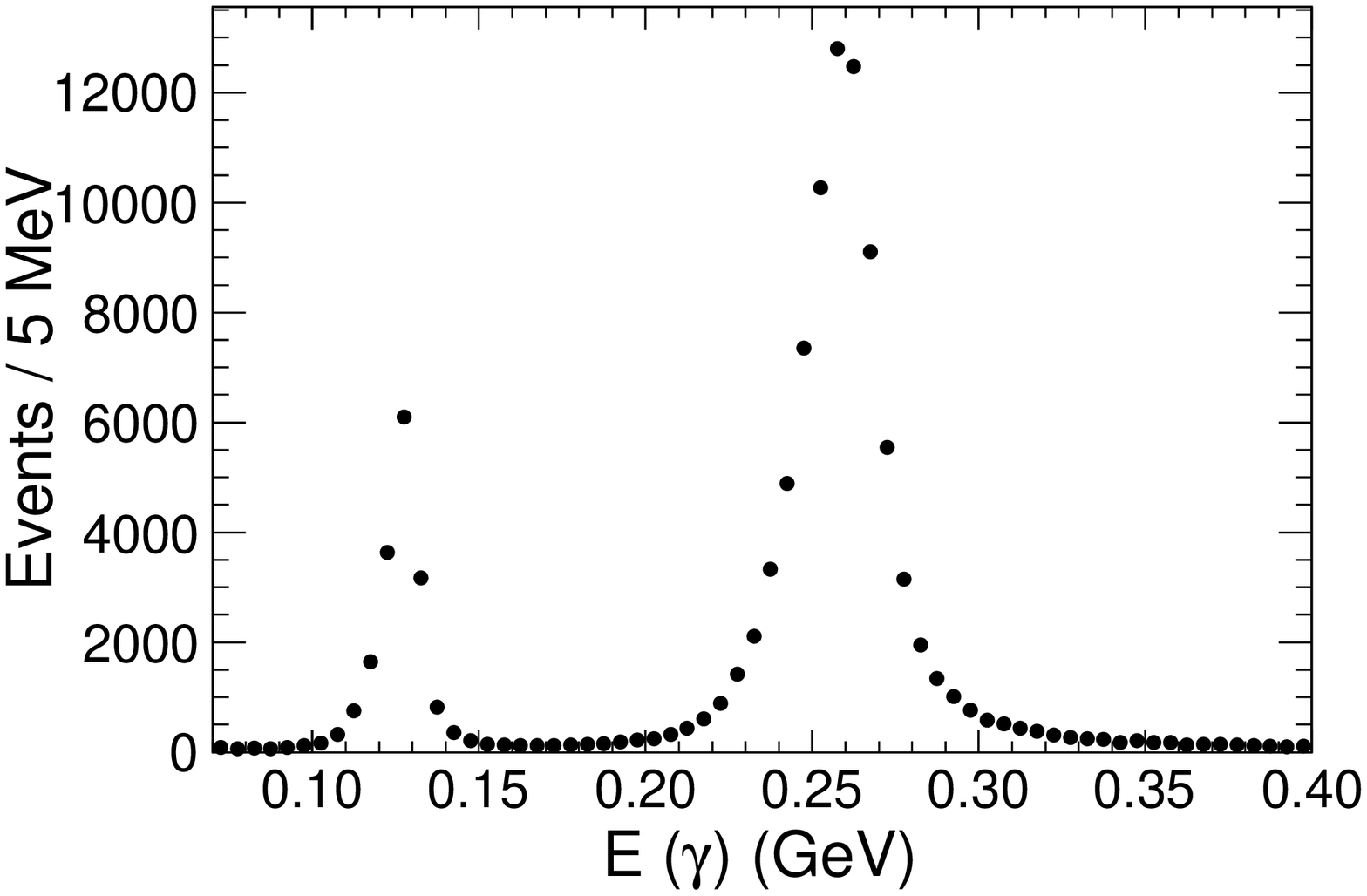}
   \end{overpic}
  \vskip -0.8cm
  \hskip 0.5cm
   \caption{The $E(\gamma)$ spectrum for the radiative photon in the samples $\psip\to\gamma\chicj,\chicj\to\kpkm$. } \label{KK}
  \end{figure}

An unbinned maximum likelihood~(ML) fit is performed to the $E(\gamma_1)$ spectrum as shown in Fig.~\ref{FIT}  to extract the signal yields. In the fit, the non-peaking background is described with the function:
 \begin{equation}\label{BKG}
f_{bg}= p_0 + p_1 E + p_2 E^2 + p_3 E^a,
\end{equation}
where $p_0$, $p_1$, $p_2$, $p_3$ and $a$ are free parameters and are determined in the fit. The reliability of the background function is validated using  the $\psipp$ data sample taken at $\sqrt{s} = 3.773$~GeV  and the off-resonance data sample taken at $\sqrt{s}=3.65$~GeV. Figure~\ref{background} shows the corresponding $E(\gamma_1)$ spectrum for the $\psipp$ data sample (upper plot) and the off-resonance data sample (lower plot), where the transition to either  $\chicz$ or $\chict$ in $\psipp$ data sample is expected to be less than 12.9 events~\cite{BESChiCJ} and can be neglected. As shown in Fig.~\ref{background}, we fit the $E(\gamma_1)$ distribution of the $\psipp$ data sample with the Eq.~(\ref{BKG}) and obtain an excellent agreement between the data and fit curve.  We also plot the $E(\gamma_1)$ distributions of the $\psipp$ data sample overlaid with the $E(\gamma_1)$ distributions of the off-resonance data sample, normalized to the same luminosity, and a good agreement is also obtained. The shapes of the $\chicz$ and $\chict$ resonances used in the fit are modeled with a nearly background-free control sample $\psip\to\gamma\chicj,\chicj\to\kpkm$. The MC studies indicate that the control sample has similar resolution on $E(\gamma_1)$ distribution to that of interest. The purity of the control sample is larger than 99.5\%, and the corresponding $E(\gamma_1)$ spectrum is shown in Fig.~\ref{KK}. In the fit, the shapes of $\chicj$ signal are fixed accordingly and the yields are free parameters.

The resultant signal yields are $N(\chicz)=3542.0 \pm 139.4$ and $N(\chict)=5044.9 \pm 138.3$, after subtraction the peaking backgrounds listed in Table~\ref{ExMC01}. The product of the branching fractions is determined by:
\begin{equation*}
\mathcal{B}(\psip\to\gamma\chicJ)\cdot\mathcal{B}(\chicJ \to \GG)  = \frac{N(\chicJ)}{N_{\psip}\cdot\epsilon(\chicJ)},
\end{equation*}
where $N_{\psip}$ is the total number of $\psip$. By incorporating the decay branching fraction $\psip\to\gamma\chicJ$ and the total width of $\chicj$ from the PDG average values:
\begin{equation}
\begin{split}
\mathcal{B}(\psip \to \gamma\chicz) & = (9.99 \pm 0.27)\%, \\
\Gamma(\chicz) & = (10.5 \pm 0.6) ~\rm{MeV}, \\
\mathcal{B}(\psip \to \gamma\chict) & = (9.11 \pm 0.31)\%, \\
\Gamma(\chict) & = (1.93 \pm 0.11) ~\rm{MeV}, \\
\end{split}
\end{equation}
we further determine $\chicj$ two-photon decay branching fraction $\mathcal{B}(\chicj\to\GG)$, the corresponding partial decay width $\Gamma_{\GG}(\chicj)$, as well as the ratio of the two measured partial decay width $\mathcal{R}$. All of the above numerical results are summarized in Table~\ref{Branching}.

\begin{table*}[htp]
\begin{center}
\parbox[2cm]{0.95\textwidth}{
\caption{Summary of the measurement. The first uncertainty is statistical, second is systematic and third is from the uncertainties associated with the branching fraction of $\psip\to\gamma \chicj$, and the total decay width of $\chicj$ quoted from PDG. The common systematic uncertainties, which are described in Table~\ref{systematic}, have been canceled in determining $\mathcal{R}$. Here, $\mathcal{B}_1 \equiv \mathcal{B}(\psip\to\gamma\chicj)$, $ \mathcal{B}_2 \equiv \mathcal{B}(\chicj\to\GG)$, $\Gamma_{\GG}\equiv\Gamma_{\GG}(\chicj\to\GG)$, and $\mathcal{R}\equiv \Gamma_{\GG}(\chict)/\Gamma_{\GG}(\chicz)$. }\label{Branching}}
\vspace{0.2cm}
\begin {tabular*}{0.95\textwidth}{@{\extracolsep{\fill}}lcc}
\hline\hline
Quantity & $\chicz$ & $\chict$ \\
\hline
$\mathcal{B}_1\times\mathcal{B}_2~(10^{-5})$ & $1.93 \pm 0.08 \pm 0.05$ & $2.83 \pm 0.08 \pm 0.06$ \\
$\mathcal{B}_2~(10^{-4})$ & $1.93 \pm 0.08 \pm 0.05 \pm 0.05$ & $3.10 \pm 0.09 \pm 0.07 \pm 0.11$ \\
$\Gamma_{\GG}(\rm{keV})$ & $2.03 \pm 0.08 \pm 0.06 \pm 0.13$ & $0.60 \pm 0.02 \pm 0.01 \pm 0.04$ \\
$\mathcal{R}$ & \multicolumn{2}{c}{$0.295 \pm 0.014 \pm 0.007 \pm 0.027$} \\
\hline\hline
\end{tabular*}
\end{center}
\end{table*}

Several systematic uncertainties in the measurement of the branching fractions are considered, including those associated with the total number of $\psip$ events, the photon detection and reconstruction efficiency, the kinematic fit, the fitting procedure and peaking background subtraction. Most systematic uncertainties are determined by comparing the behavior between the MC simulation and data for certain very clean and high-statistics samples.

\begin{table}[htp]
\begin{center}
\caption{Summary of the systematic uncertainties~(in \%).}\label{systematic}
\vspace{0.2cm}
\begin {tabular}{lcc}
\hline\hline
Sources & $\chicz$ & $\chict$ \\
\hline
Number of $\psip$  & \mulC{2}{c}{0.7}  \\
Photon Detection &  \mulC{2}{c}{1.5}  \\
Kinematic Fit &  \mulC{2}{c}{1.0} \\
Neutral Trigger Efficiency  &  \mulC{2}{c}{0.1} \\
Fit Procedure & 2.0 & 1.2 \\
Peaking Background & 0.3 & 0.1 \\
Helicity Two Assumption & $-$ & 0.2 \\
\hline
Total & 2.8 & 2.3 \\
\hline\hline
\end{tabular}
\end{center}
\end{table}

The number of $\psip$ events, $N_{\psip}$, is determined by analyzing the inclusive hadronic events with the procedure described in detail in Ref.~\cite{NpsipAll}. The uncertainty of the total number of $\psip$ events is 0.7\%.

The three photons in the final states include a soft photon from the radiative transition and two high-energetic photons from $\chicj$ decays. The photon detection efficiency and its uncertainty for low energy photons are studied using three different methods described in Ref.~\cite{Photon01}. On average, the efficiency difference between data and MC simulation is less than 1\%. The average momenta of the two high-energy photons are about 1.7 $\rm{GeV}/c$. The corresponding systematic uncertainty on its reconstruction is determined to be 0.25\% per photon as described in Ref.~\cite{Photon02}, which is estimated based on a control sample of $\jpsi\to\gamma\eta^{\prime}$. The total uncertainty associated with the reconstruction of the three photons is 1.5\%.

To suppress the background, the number of selected photon candidates is required to be exactly three. An alternative analysis is performed by requiring at least three photons.
Looping over all the three photon combinations in the 4C kinematic fit, we take the combination with the minimum $\chi^2$ for this fit as the final photon candidates. We then perform the same procedure to extract the final results, and the resultant changes with respect to the nominal values are found to be very small. Thus the uncertainty associated with the requirement of exactly three photons is negligible.

The uncertainty due to the kinematic fit is estimated using a sample of $\EE\to\GGG$, which has the same event topology as the signal. We select the sample by using off-resonance data taken at $\sqrt{s}=3.65$ GeV to determine the efficiency difference between data and MC simulation for the requirement of $\chi^2_{4C}<80$ in the 4C fit, where the efficiency of the 4C kinematic fit is the ratio of the number of the events with and without the 4C fit. The uncertainty due to the kinematic fit is determined to be 1.0\%.

The signal shapes are obtained from $\EE\to\gamma\chicj, \chicj\to\kpkm$ events in the data. Considering the resolutions differ slightly between $\EE\to\gamma\chicj, \chicj\to\gamma\gamma$ and $\chicj\to\kpkm$, the uncertainty due to the signal shape is estimated by the alternative fit using signal MC shapes instead. The shape of the continuum background is parameterized using Eq.~(\ref{BKG}). The systematic uncertainty due to the choice of parameterization for the background shape is estimated by varying the fitting range and the order of the polynomial. The relative changes on the $\chicz$ and $\chict$ signal yields, 2.0\% and 1.2\%, respectively, are taken as the uncertainties associated with the fit procedure.

The expected number of peaking background events from $\chicj\to\piz\piz(\eta\eta)$ decays, summarized in Table~\ref{ExMC01}, are subtracted from the fit results. We change the number of peaking background by one standard deviation of the uncertainties when recalculating the signal yields.  The resultant changes on signal yields, 0.3\% and 0.1\% for $\chicz$ and $\chict$, respectively, are taken as the uncertainties.

The systematic uncertainty due to the trigger efficiency in these neutral channels is estimated to be smaller than 0.1\%, based on cross-checks using different trigger conditions~\cite{Trigger}.

While generating MC samples, we assume a pure helicity-two decay of $\chict\to\GG$. In a relativistic calculation, Barnes~\cite{Pre02} predicted the helicity-zero component to be about 0.5\%. In Sec.~\ref{Hilicity}, the ratio of the two photon widths for the helicity-zero and helicity-two amplitudes is measured to be $(0.0 \pm 0.6 \pm 1.2)\times10^{-2}$.  By including a helicity-zero fraction of 2\% in the MC samples, we conservatively estimate the uncertainty associated with the helicity-zero component to be 0.2\%

All of the above systematic uncertainties are listed in Table~\ref{systematic}.  
We assume that all systematic uncertainties are independent and add them in quadrature to obtain the total systematic uncertainty (except for the ratio $\mathcal{R}$, where the first four contributions in Table~\ref{systematic} cancel). For the calculations of the branching fraction $\mathcal{B}(\chicj\to\GG)$ and the corresponding two-photon partial decay widths $\Gamma_{\GG}(\chicj)$, the uncertainties related with the branching fractions $\mathcal{B}(\psip\to\gamma\chicj)$ and the full decay widths $\Gamma(\chicj)$ are quoted separately as the second systematic uncertainty.

By including an additional resonance corresponding to $\chico$ in the fit to the $E(\gamma_1)$ spectrum of Fig.~\ref{FIT} , we examine the existence of the decay $\chico\to\GG$, which is forbidden by the Landau-Yang theorem. The shape of the $\chico$ signal is parameterized using a smoothed MC histogram convolved with a Gaussian function, $G(0,\sigma)$, where $\sigma$ is fixed to the resolution difference between data and MC simulation of the $\chicz\to\kpkm$ process. The efficiency is $(39.80 \pm 0.04)\%$. The systematic uncertainties are similar to $\chicz\to\GG$, except for the uncertainties from peaking background subtraction and from the branching fraction of $\psip\to\gamma \chico$ quoted from PDG. The likelihood function is determined as a function of the branching fraction $\mathcal{B}(\chico\to\GG)$. The corresponding  systematic uncertainty in the branching fraction measurement is incorporated by convolving the likelihood function with a Gaussian function, where the width of Gaussian function is the total systematic uncertainty. Incorporating the decay branching fraction $\mathcal{B}(\psip\to\gamma\chico) = (9.55 \pm 0.31)\%$ and the total decay width $\Gamma(\chico) = (0.84 \pm 0.04)~\rm{MeV}$ quoted from the PDG~\cite{PDG}, we obtain the upper limit at the 90\% confidence level for the branching fraction $\mathcal{B}(\chico\to\GG) < 6.3\times 10^{-6}$ and for the two-photon partial decay width $\Gamma_{\GG}(\chico)<5.3~\rm{eV}$, which are much more stringent than those of previous measurements.

\section{\boldmath Helicity Amplitude Analysis For $\chict\to\GG$ }\label{Hilicity}

In the $\chict\to\GG$ decay, the final state is a superposition of helicity-zero~($\lambda=0$) and helicity-two~($\lambda=2$) components, where $\lambda$ is the difference of helicity between the two photons. The formulae for the helicity amplitudes in $\psip \to \gamma_1\chict, \chict \to \gamma_2\gamma_3$, including high-order multipole amplitudes, is shown in Eq.~(\ref{Chic2}):
\begin{widetext}
\begin{equation}\label{Chic2}
\begin{split}
W_2(\theta_1,\theta_2,\phi_2)
& = f_{0/2}\Big[3x^2\sin^2\theta_1\sin^2\theta_2 +  \frac{3}{2}y^2(1+\cos^2\theta_1)\sin^4\theta_2\\
& - \frac{3\sqrt{2}}{2}xy\sin2\theta_1\sin^2\theta_2\sin2\theta_2\cos\phi_2 + \sqrt{3}x\sin2\theta_1\sin2\theta_2 (3\cos^2\theta_2 -1)\cos\phi_2\\
& + \sqrt{6}y\sin^2\theta_1\sin^2\theta_2 (3\cos^2\theta_2 -1)\cos2\phi_2 + (1+\cos^2\theta_1)(3\cos^2\theta_2 -1)^2\Big]_{\lambda=0} \\
& + \Big [ 2x^2\sin^2\theta_1(1+\cos^2\theta_2)\sin^2\theta_2 + \frac{1}{4}y^2(1+\cos^2\theta_1)(1+6\cos^2\theta_2+\cos^4\theta_2)\\
& + \frac{\sqrt{2}}{4}xy\sin2\theta_1\sin^2\theta_2(3+\cos^2\theta_2)\cos\phi_2 - \frac{\sqrt{3}}{2}x\sin2\theta_1\sin2\theta_2\sin^2\theta_2\cos\phi_2\\
& + \frac{\sqrt{6}}{2}y\sin^2\theta_1(1-\cos^4\theta_2)\cos2\phi_2 + \frac{3}{2}(1+\cos^2\theta_1)\sin^4\theta_2\Big ]_{\lambda=2} \; , \\
\end{split}
\end{equation}
 \end{widetext}
where $x=A_1/A_0$, $y=A_2/A_0$, and $A_{0,1,2}$ are the amplitude of $\chict$ production with helicity 0, 1, 2, respectively. $\theta_1$ is the polar angle of the radiative photon, with respect to the direction of the positron beam, $\theta_2$ and $\phi_2$ are the polar angle and azimuthal angle of one of the photons in the decay $\chict\to\GG$ in the $\chict$ rest frame, relative to the radiative photon direction as polar axis, and $\phi_2 = 0$ is defined by the electron beam direction. The quantity $f_{0/2} = |F_0|^2/|F_2|^2$ is the ratio of partial two-photon decay widths between the helicity-zero and helicity-two components, where $F_0 (F_2)$ is the decay amplitude of the helicity $ \lambda = 0 (2)$ component.

An unbinned ML fit to the angular distribution is performed to the candidate of $\chict\to\GG$ to determine $x$, $y$ and $f_{0/2}$. For convenience, we define 12 new factors, $a_1, a_2, ..., a_{12}$, which are:
\begin{equation}
a_1 = 3\sin^2\theta_1\sin^2\theta_2,
\end{equation}
\begin{equation}
a_2 = \frac{3}{2}(1+\cos^2\theta_1)\sin^4\theta_2,
\end{equation}
\begin{equation}
a_3 =  - \frac{3\sqrt{2}}{2}\sin2\theta_1\sin^2\theta_2\sin2\theta_2\cos\phi_2,
\end{equation}
\begin{equation}
a_4 = \sqrt{3}\sin2\theta_1\sin2\theta_2 (3\cos^2\theta_2 -1)\cos\phi_2,
\end{equation}
\begin{equation}
a_5 = \sqrt{6}\sin^2\theta_1\sin^2\theta_2 (3\cos^2\theta_2 -1)\cos2\phi_2,
\end{equation}
\begin{equation}
a_6 = (1+\cos^2\theta_1)(3\cos^2\theta_2 -1)^2,
\end{equation}
\begin{equation}
a_7 = 2\sin^2\theta_1(1+\cos^2\theta_2)\sin^2\theta_2,
\end{equation}
\begin{equation}
a_8 = \frac{1}{4}(1+\cos^2\theta_1)(1+6\cos^2\theta_2+\cos^4\theta_2),
\end{equation}
\begin{equation}
a_9 = \frac{\sqrt{2}}{4}\sin2\theta_1\sin^2\theta_2(3+\cos^2\theta_2)\cos\phi_2,
\end{equation}
\begin{equation}
a_{10} = -\frac{\sqrt{3}}{2}\sin2\theta_1\sin2\theta_2\sin^2\theta_2\cos\phi_2,
\end{equation}
\begin{equation}
a_{11} = \frac{\sqrt{6}}{2}\sin^2\theta_1(1-\cos^4\theta_2)\cos2\phi_2,
\end{equation}
\begin{equation}
a_{12} = \frac{3}{2}(1+\cos^2\theta_1)\sin^4\theta_2.
\end{equation}

To obtain a normalized decay amplitude by considering the detection acceptance and efficiency effects, we calculate the average values of $a_n$ with the MC sample  of $\psip\to\gamma\chict,\chict\to\GG$ generated with a uniform distribution in phase space:
\begin{equation}
\bar{a}_n = \frac{\sum^N_{i=1}a_n(i)}{N}, n = 1, 2, ..., 12,
\end{equation}
where $N$ is the number of MC events after applying all the selection criteria.

The normalized probability density function is written as:
\begin{widetext}
\begin{equation}
f(x,y,f_{0/2}) = \frac{W_2(\theta_1,\theta_2,\phi_2|x,y,f_{0/2})}{f_{0/2}(\bar{a}_1x^2+\bar{a}_2y^2+\bar{a}_3xy+\bar{a}_4x+\bar{a}_5y+\bar{a}_6) + \bar{a}_7x^2+\bar{a}_8y^2+\bar{a}_9xy+\bar{a}_{10}x+\bar{a}_{11}y+\bar{a}_{12}}.
\end{equation}
\end{widetext}
A joint likelihood function is constructed as $\ln\mathcal{L} = \sum^n_{i=1} \ln f_i(x,y,f_{0/2})$, where the sum runs over all the events in the signal region, defined as $0.11<E(\gamma_1)<0.14$ GeV. The background contribution to the likelihood function ($\ln\mathcal{L}_b$) is evaluated with the events in the sideband regions, defined as $0.07<E(\gamma_1)<0.09$ GeV~(lower) and $0.16<E(\gamma_1)<0.19$ GeV~(upper) and normalized according to the numbers of background events in the signal and sideband regions evaluated with the fit results to the $E(\gamma_1)$ distribution. We maximize the function $\ln\mathcal{L}_s = \ln\mathcal{L} - \ln\mathcal{L}_b$ to extract best values of $x$, $y$ and $f_{0/2}$. 

\begin{figure}[htbp]
   \centering
  \begin{overpic}[width=7.5cm,height=9.0cm,angle=0]{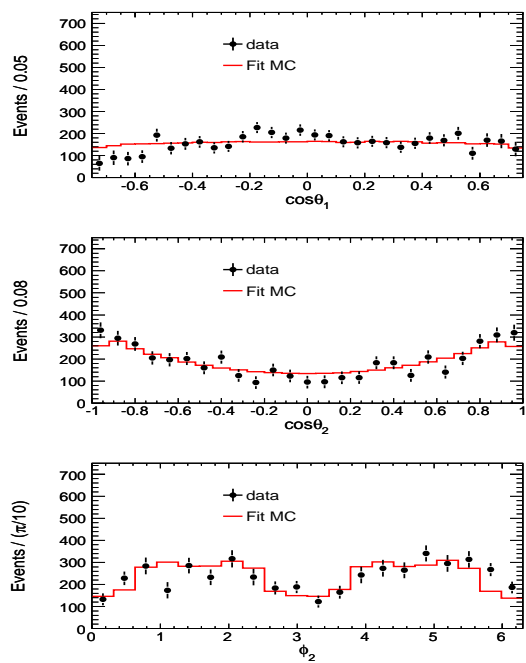}
   \end{overpic}
  \vskip -0.8cm
  \hskip 0.5cm
   \caption{Distribution of $\cos\theta_1$, $\cos\theta_2$ and $\phi_2$ for the decay $\psip\to\gamma\chict,\chict\to\GG$, where the dots with error bar indicate background-subtracted data and the histograms show the fitted results. } \label{Angular}
  \end{figure}

In the nominal fit, the values for $x$ and $y$ are fixed to the values~($x=1.55$ and $y=2.10$) obtained in Ref.~\cite{ChiC2} from a sample of 13800 $\psip\to\gamma\chict,\chict\to K^+K^-, \pi^+\pi^-$ events. The remaining parameter $f_{0/2}$ is determined to be:
\begin{equation}
f_{0/2} =  (0.0 \pm 0.6)\times 10^{-2},
\end{equation}
where the uncertainty is statistical only from the fit. The angular distributions of background-subtracted data and the fit results are shown in Fig.~\ref{Angular}, where the fit curves are produced from the MC events generated incorporating the angular distribution (Eq.~(\ref{Chic2})) with the parameters $x=1.55$, $y=2.10$, $f_{0/2}=0.0$. It is found that the angular distributions are consistent between the data and the fit curves within the statistical uncertainty.

The goodness of the fit is estimated using the Pearson-$\chi^2$ test. The data and MC simulation are divided into 8 bins with identical size in each dimensional~($\cos\theta_1$, $\cos\theta_2$, $\phi_2$) of the three-dimension angular distribution, for a total of $8^3$ cells. The $\chi^2$ is defined as:
\begin{equation}
\chi^2 = \sum_i\frac{(n_i^\text{DT}-n_i^\text{MC})^2}{\sigma^2_{n_i^\text{DT}}},
\end{equation}
where $n_i^\text{DT} (\sigma_{n_i^\text{DT}})$ is the observed number~(its statistical uncertainty) of signal events after background subtraction in the $i^{th}$ bin from data and $n_i^\text{MC}$ is the expected number of events predicted from MC simulation according to the fit results. If the number of events in a bin is less than 5, the events are merged with an adjacent bin. The resultant $\chi^2$ of test is $\chi^2/\rm{ndf} = 1.04$, indicates an reasonable fit quality, where ndf is the number of degrees of freedom.

An alternative fit to the data with free parameters $x$ and $y$ is performed to test the reliability of the fit.  This fit returns
\begin{equation*}
x = 1.68 \pm 0.11, y = 2.21 \pm 0.13,\\
f_{0/2} = (0.0 \pm 0.7)\times 10^{-2},
\end{equation*}
where the uncertainties are statistical only. The values for $x, y$ are consistent with the more precise results in Ref.~\cite{ChiC2} and that for $f_{0/2}$ is consistent with our nominal analysis.  

In the measurement of the amplitude ratio between different helicity components, $f_{0/2}$, many systematic uncertainties cancel. Only the effects due to the inconsistency between data and MC simulation dependence on the polar angle, the uncertainties of the input $x$ and $y$ parameters, background subtraction and $\chicz$ contamination are considered.

As discussed above, in the nominal fit, the parameters $x$ and $y$ are fixed to the measured values from Ref.~\cite{ChiC2}, and the ratio $f_{0/2}$ is determined. We change the input $x$ and $y$ values by one standard deviation of their uncertainties and repeat the fit. To estimate the uncertainty due to background subtraction, we repeat the fit by varying the sideband regions from~(0.07,~0.09) GeV~(lower) and~(0.16,~0.19) GeV~(upper) to~(0.07,~0.10) GeV and~(0.15,~0.19) GeV. The resultant changes on $f_{0/2}$ with respect to the nominal value in the above two cases are found to be negligible. From MC simulations, we find that only 0.044\% of the $\chicz\to\GG$  events enter the $\chict\to\GG$ signal region, and thus any related uncertainty is ignored.

The uncertainty due to the polar-angle dependent inconsistency between data and MC simulation is estimated using $\chicz$ events. The inconsistency consists of the discrepancy associated with the energy resolution and detection efficiency for photon, the kinematic fit, the trigger efficiency, selection efficiency, and the method to subtract the background. The reliability of this method has been validated by many analyses~\cite{LUOT,ChiC2,BESII}. Since the $\chicz$ is pure helicity-zero, the $x$ and $y$ parameters in Eq.~(\ref{Chic2}) are expected to be zero. For the $\chicz\to\GG$ decay, the helicity value difference between the two photons is also expected to be zero, which means only the $\lambda=0$ term in Eq.~(\ref{Chic2}) remains. Accordingly we modify Eq.~(\ref{Chic2}) to:
\begin{widetext}
\begin{equation}\label{Chic0}
\begin{split}
W_0(\theta_1,\theta_2,\phi_2)
& = \Big[3x^2\sin^2\theta_1\sin^2\theta_2 + \frac{3}{2}y^2(1+\cos^2\theta_1)\sin^4\theta_2 \\
& - \frac{3\sqrt{2}}{2}xy\sin2\theta_1\sin^2\theta_2\sin2\theta_2\cos\phi_2 + \sqrt{3}x\sin2\theta_1\sin2\theta_2 (3\cos^2\theta_2 -1)\cos\phi_2\\
& + \sqrt{6}y\sin^2\theta_1\sin^2\theta_2 (3\cos^2\theta_2 -1)\cos2\phi_2 + (1+\cos^2\theta_1)\Big]_{\lambda=0} \\
& + f_{2/0}\Big[2x^2\sin^2\theta_1(1+\cos^2\theta_2)\sin^2\theta_2 + \frac{1}{4}y^2(1+\cos^2\theta_1)(1+6\cos^2\theta_2+\cos^4\theta_2) \\
& + \frac{\sqrt{2}}{4}xy\sin2\theta_1\sin^2\theta_2(3+\cos^2\theta_2)\cos\phi_2 - \frac{\sqrt{3}}{2}x\sin2\theta_1\sin2\theta_2\sin^2\theta_2\cos\phi_2\\
& + \frac{\sqrt{6}}{2}y\sin^2\theta_1(1-\cos^4\theta_2)\cos2\phi_2 + \frac{3}{2}(1+\cos^2\theta_1)\sin^4\theta_2\Big]_{\lambda=2} \; . \\
\end{split}
\end{equation}
\end{widetext}
We then fit the events in $\chicz$ signal region with Eq.~(\ref{Chic0}) by a similar method as applied to the $\chict$ signal. Non-zero $x$, $y$ or $f_{2/0}$ values indicate the inconsistency between data and MC simulation. To be conservative, the sum of any shift from 0 plus its uncertainty will be taken as the net systematic effect. The fitted result is $f_{2/0}=0.000 \pm 0.012$ when $x$ and $y$ are fixed to be zero. Studies with MC samples demonstrate that a systematic uncertainty in modeling the $\theta_1$, $\theta_2$ and $\phi_2$ efficiency produces a shift of approximately the same size for $f_{2/0}$ in $\chicz$ sample and $f_{0/2}$ in $\chict$ sample. Therefore, the observed shift from $f_{2/0}$ for the $\chicz$ data can be used to estimate the corresponding systematic uncertainty in the $\chict\to\GG$ measurement. Thus we take 0.012 as the systematic uncertainty.

\section{Conclusion}

In summary, we present the updated measurements of the two-photon decays of $\chicj$ via the radiative transition $\psip \to \gamma\chicj$ based on a $\psip$ data sample of $448.1\times10^6$ events. We determine $\mathcal{B}(\chicz\to\GG)  = (1.93 \pm 0.08 \pm 0.05 \pm 0.05)\times 10^{-4}$ and $\mathcal{B}(\chict\to\GG) = (3.10 \pm 0.09 \pm 0.07 \pm 0.11)\times 10^{-4} $, which agree with the previous measurements~\cite{LUOT,CLEOc}. Incorporating the branching fraction $\mathcal{B}(\psip\to\gamma\chicj)$ and the total decay widths $\Gamma(\chicj)$ quoted from PDG, we also determine the decay branching fractions and the two-photon partial decay widths of $\chicj\to\gamma\gamma$, as well as the ratio of two-photons partial decay width between $\chict$ and $\chicz$. A comparison between this measurement, the previous measurements, and the PDG world average values is summarized in Table~\ref{comparison}; our results are the most precise to date.

\begin{widetext}
\begin{threeparttable}[htp]
\begin{center}
\begin{small}
\caption{The comparison of experimental results for the two-photon partial widths of $\chicz$ and $\chict$.  }\label{comparison}
\vspace{0.2cm}
\begin {tabular*}{0.95\textwidth}{@{\extracolsep{\fill}}lcccc}
\hline\hline
Quantity & PDG average values\tnote{a} & CLEO-c\tnote{b} & BESIII\tnote{b} & This measurement\tnote{b}  \\
\hline
$\mathcal{B}_1\times\mathcal{B}_2(10^{-5})(\chicz)$\tnote{c}  & $2.23\pm0.14$ & $2.17 \pm 0.32 \pm 0.10$ & $2.17 \pm 0.17 \pm 0.12$ &  $1.93 \pm 0.08 \pm 0.05$  \\
$\mathcal{B}_1\times\mathcal{B}_2(10^{-5})(\chict)$\tnote{c} & $2.50\pm0.15$  & $2.68 \pm 0.28 \pm 0.15$ & $2.81 \pm 0.17 \pm 0.15$ & $2.83 \pm 0.08 \pm 0.06$ \\
$\mathcal{B}_2(10^{-4})(\chicz)$\tnote{c}  & $2.23 \pm 0.13 $ & $2.31 \pm 0.34 \pm 0.15$ & $2.24 \pm 0.19 \pm 0.15$ & $1.93 \pm 0.08 \pm 0.07$ \\
$\mathcal{B}_2(10^{-4})(\chict)$\tnote{c}  & $2.74 \pm 0.14 $ &  $3.23 \pm 0.34 \pm 0.24$  & $3.21 \pm 0.18 \pm 0.22$ & $3.10 \pm 0.09 \pm 0.13$\\
$\Gamma_{\GG}(\chicz)~\rm{keV}$ & $2.24 \pm 0.19 $ &  $2.36 \pm 0.35 \pm 0.22$  & $2.33 \pm 0.20 \pm 0.22$  &$2.03 \pm 0.08 \pm 0.14$ \\
$\Gamma_{\GG}(\chict)~\rm{keV}$ & $0.53 \pm 0.03 $ &  $0.66 \pm 0.07 \pm 0.06$  & $0.63 \pm 0.04 \pm 0.06$  &$0.60 \pm 0.02 \pm 0.04$ \\
$\mathcal{R}$  & $0.236 \pm 0.024 $ & $0.28 \pm 0.05 \pm 0.04$  & $0.271 \pm 0.029 \pm 0.030 $ &  $0.295 \pm 0.014 \pm 0.028$ \\
$f_{0/2}(10^{-2})$ & ... & ... & $0\pm 2\pm 2$ & $0.0\pm0.6\pm1.2$\\
\hline\hline
\end{tabular*}
\begin{tablenotes}
  \item[a] The results from the literature have been reevaluated by using the branching fractions and the total width from PDG.
  \item[b] The first uncertainty is statistical, the second is systematic uncertainty including those from branching fraction $\mathcal{B}(\psip\to\gamma\chicj)$ and the total decay widths $\Gamma(\chicj)$.
  \item[c] $\mathcal{B}_1 \equiv \mathcal{B}(\psip\to\gamma\chi_{c0,2})$, $ \mathcal{B}_2 \equiv \mathcal{B}(\chicj\to\GG)$, $\Gamma_{\GG}(\chicj)\equiv\Gamma_{\GG}(\chicj\to\GG)$, $\mathcal{R}\equiv \Gamma_{\GG}(\chict)/\Gamma_{\GG}(\chicz)$.
\end{tablenotes}
\end{small}
\end{center}
\end{threeparttable}
\end{widetext}

We also search for the decay $\chico\to\gamma\gamma$, which is forbidden by the Landau-Yang theorem, by examining the $E_{\gamma}$ distribution. We do not find an obvious $\chico\to\gamma\gamma$ signal, and an upper limit at the 90\% confidence level on the decay branching fractions and the two-photon partial width are set to be $\mathcal{B}(\chico\to\GG)<6.3\times10^{-6}$ and $\Gamma_{\GG}(\chico)<5.3~\rm{eV}$, respectively.

The ratio of two-photon partial decay widths between $\chict$ and $\chicz$ is measured to be $\mathcal{R}=0.295\pm0.014\pm0.007\pm0.027$. This is larger than the theoretical calculation taking into consideration the first order radiative correction~\cite{MBV}, which obtains a reduction from the nominal $4/15 = 0.267$ by a multiplicative factor of $(1-5.51\alpha_s/\pi)$. This may indicate an inadequacy of the calculation; higher-order radiative correction calculations are desirable. Alternatively, as noted by Buchm\"{u}ller~\cite{Buch}, a different scheme or scale of the the renormalization is necessary to obtain better convergence for the radiative corrections. Moreover, the precise $\mathcal{R}$ values obtained can help to calibrate the different theoretical potential models~\cite{Pre01,Pre02,Pre03,Pre04,Pre05,Pre06,Pre07,Pre11,Pre08}.

Additionally, we also perform a helicity amplitude analysis for the decay of $\psip\to\gamma\chict,\chict\to\GG$. The ratio of the two-photon partial widths between the helicity-zero and helicity-two components in the decay of $\chict\to\GG$ is determined to be $f_{0/2}=(0.0\pm0.6\pm1.2)\times10^{-2}$, confirming that helicity-zero component is highly suppressed. This more precise measurement is consistent with the previous experimental results~\cite{LUOT}.

\section{Acknowledgments}

The BESIII collaboration thanks the staff of BEPCII and the IHEP computing center for their strong support. This work is supported in part by National Key Basic Research Program of China under Contract No. 2015CB856700; National Natural Science Foundation of China~(NSFC) under Contracts Nos. 11235011, 11322544, 11335008, 11425524, 11635010; the Chinese Academy of Sciences~(CAS) Large-Scale Scientific Facility Program; the CAS Center for Excellence in Particle Physics~(CCEPP); the Collaborative Innovation Center for Particles and Interactions~(CICPI); Joint Large-Scale Scientific Facility Funds of the NSFC and CAS under Contracts Nos. U1232201, U1332201, U1532257, U1532258; CAS under Contracts Nos. KJCX2-YW-N29, KJCX2-YW-N45; 100 Talents Program of CAS; National 1000 Talents Program of China; INPAC and Shanghai Key Laboratory for Particle Physics and Cosmology; German Research Foundation DFG under Contracts Nos. Collaborative Research Center CRC 1044, FOR 2359; Istituto Nazionale di Fisica Nucleare, Italy; Koninklijke Nederlandse Akademie van Wetenschappen~(KNAW) under Contract No. 530-4CDP03; Ministry of Development of Turkey under Contract No. DPT2006K-120470; National Science and Technology fund; The Swedish Resarch Council; U. S. Department of Energy under Contracts Nos. DE-FG02-05ER41374, DE-SC-0010118, DE-SC-0010504, DE-SC-0012069; University of Groningen~(RuG) and the Helmholtzzentrum fuer Schwerionenforschung GmbH~(GSI), Darmstadt; WCU Program of National Research Foundation of Korea under Contract No. R32-2008-000-10155-0.

\end{document}